\documentclass{kapproc} 

\RequirePackage{graphicx}%
\RequirePackage{epsf}%
\input{psfig}

\upperandlowercase
\let\footnote\savefootnote
\let\footnotetext\savefootnotetext 
 
\kluwerbib 

\newcommand{\etal}{\emph{et al.}\ }
\newcommand{\eg}{\emph{e.g.}\ }

\newcommand{\msun}{M$_\odot$}

\begin{document}


\articletitle{Detecting Primordial Stars}


\chaptitlerunninghead{Detecting Primordial Stars}



 \author{Nino Panagia}
 \affil{ESA/Space Telescope Science Institute, 3700 San Martin Drive,
 Baltimore, MD, USA}
 \email{panagia@stsci.edu}





 \begin{abstract}
We discuss the expected properties  of the first stellar generations in
the Universe. We find that it is possible to discern truly primordial
populations from the next generation of stars by measuring the
metallicity of high-z star forming objects. The very low background of
the future James Webb Space Telescope ({\it JWST}) will enable it to
image and study first-light sources at very high redshifts, whereas its
relatively small collecting area limits its capability in obtaining
spectra of z$\sim$10--15 first-light sources to either the bright end
of their luminosity function or to strongly lensed sources. With a
suitable investment of observing time ~{\it JWST} will be able to
detect {\it individual} Population III supernovae, thus identifying the
very first stars that formed in the Universe.

 \end{abstract}

\section{Introduction}
\label{sec:Intro}

One of the primary goals of  modern cosmology is to answer the
question: ``When did galaxies begin to form in the early Universe and
how did they form?" Theorists predict that the formation of galaxies is
a gradual process in which progressively larger, virialized masses,
composed mostly of dark matter, harbor star formation as time elapses.
These dark-matter halos, which harbor stellar populations, then undergo a
process of hierarchical merging and evolution to become the galaxies
that make up the local Universe. In order to understand what are the
earliest building blocks of galaxies like our own, one must
detect and identify ``first light" sources, i.e, the emission from the
first objects in the Universe to undergo star formation.

The standard picture is that at zero metallicity the Jeans mass in star
forming clouds is much higher than it is in the local Universe, and,
therefore, the formation of massive stars, say, 100 \msun\/ or higher,
is highly favored. The spectral distributions of these massive stars
are characterized by effective temperatures on the Main Sequence (MS)
around $10^5$~K ({\it e.g.,} Tumlinson \& Shull 2000, Bromm {\it et
al.}~2001, Marigo {\it et al.}~2001).  Due to their temperatures these
stars are very effective in ionizing hydrogen and helium. It should be
noted that zero-metallicity (the so-called population III) stars  of
all masses have essentially the same MS luminosities as, but are much
hotter than their solar metallicity analogues.   Note also that only
stars hotter than about 90,000~K are capable of ionizing He twice in
appreciable quantities, say, more than about 10\% of the total He
content ({\it e.g., } Oliva \& Panagia 1983, Tumlinson \& Shull 2000). 
As a consequence even the most massive population III stars can produce
HeII lines only for a relatively small fraction of their lifetimes,
say, about 1~Myr or about 1/3 of their lifetimes. 

The second generation of stars forming out of pre-enriched material
will probably have different properties because cooling by metal lines
may become a viable mechanism and  stars of lower masses may be
produced (Bromm {\it et al.}\/ 2001). On the other hand, if the
metallicity is lower than about $5\times 10^{-4}$Z$_\odot$,  build up
of H$_2$ due to self-shielding may occur, thus  continuing the
formation of very massive stars (Oh \& Haiman 2002). Thus, it appears
that in the zero-metallicity case one may  expect a very top-heavy
Initial Mass Function (IMF), whereas it is not clear if the second
generation of stars is also top-heavy or follows a normal IMF. 

\section{Primordial HII Regions}

The  high effective temperatures of zero-metallicity stars imply not
only high ionizing photon fluxes for both hydrogen and helium, but
also  low optical  and UV fluxes.  
As a result,  one should
expect the rest-frame optical/UV spectrum of a primordial HII regions
to be largely dominated by its nebular emission (both continuum and
lines), so that the best strategy to detect the presence of primordial
stars is to search for the emission from associated HII regions. 

  \begin{figure}
    \begin{center}
      \includegraphics[width=4.5in]{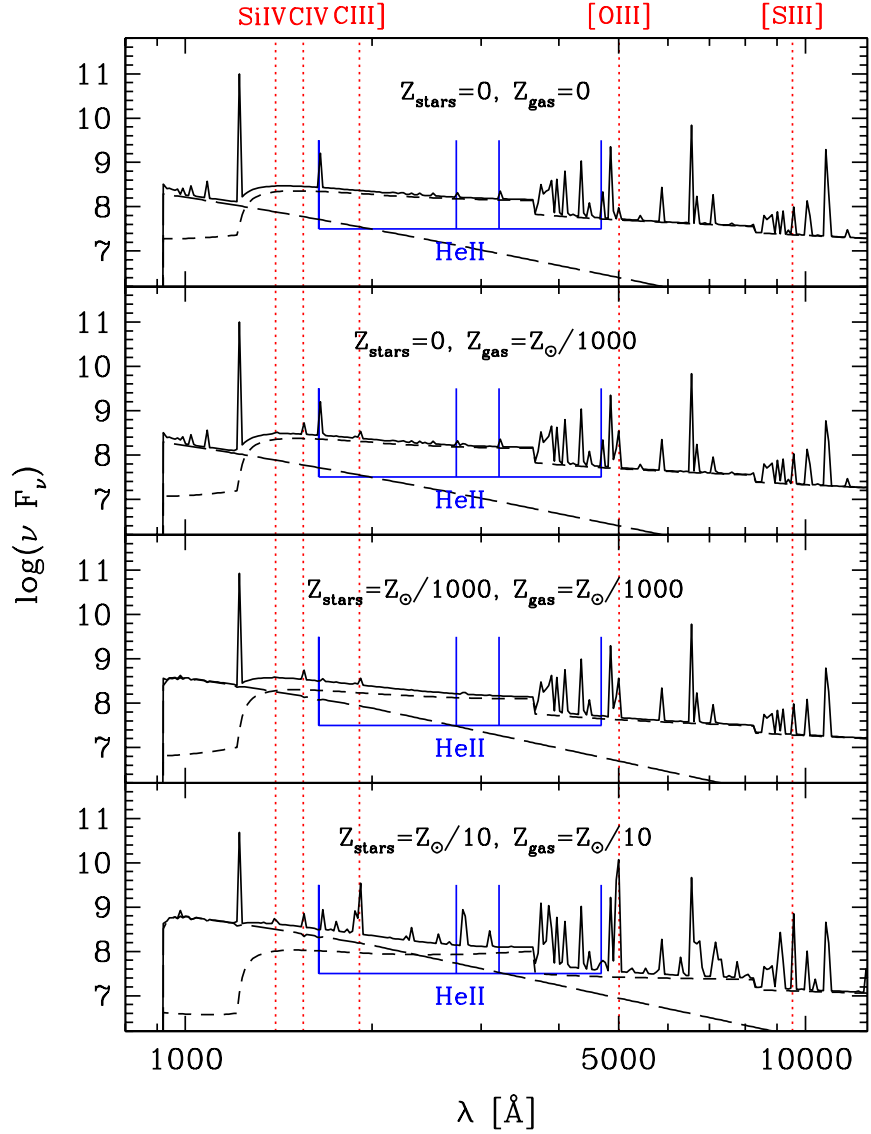}
    \end{center}
  \caption{{The synthetic spectrum of a zero-metallicity HII region
(top panel) is compared to that of HII regions with  various
combinations of stellar and nebular metallicities (lower panels). The
long-dashed and short-dashed lines represent the stellar and nebular
continua, respectively. }} 
  \end{figure}
%
%

In Panagia {\it et al.}~(2004, in preparation) we report on our
calculations of the properties of primordial, zero-metallicity HII
regions ({\it e.g.,}~ Figure~1). We find that the electron temperatures
is in excess of 20,000 K and that 45\% of the total luminosity is
converted into the Ly-$\alpha$ line, resulting in a Ly-$\alpha$
equivalent width (EW) of 3000 \AA\/  (Bromm, Kudritzki \& Loeb 2001).
The helium lines are also strong, with the HeII $\lambda$1640 intensity
comparable to that of H$\beta$ (Tumlinson {\it et al.}~2001, Panagia
{\it et al.}~2004, in preparation). An interesting feature of these
models is that the continuum longwards of Ly-$\alpha$ is dominated by
the two-photon nebular continuum. The H$\alpha$/H$\beta$ ratio for
these models is 3.2. Both the red continuum and the high
H$\alpha$/H$\beta$ ratio could be naively (and {\it incorrectly})
interpreted as a consequence of dust extinction even though no dust is
present in these systems.

From the observational point of view one will generally be unable to
measure a zero-metallicity but will usually be able to place an upper
limit to it. When would such an upper limit be indicative that one is
dealing with a population III object? According to Miralda-Escud\'e \&
Rees (1998) a metallicity Z$\simeq10^{-3}Z_\odot$ can be used as a
dividing line between the pre- and post-re-ionization Universe. A
similar value is obtained by considering that the first supernova (SN)
going off in a primordial cloud will pollute it to a metallicity of
$\sim 0.5 \times 10^{-3}Z_\odot$\ (Panagia {\it et al.}~2004, in
preparation). Thus, any object with a metallicity higher than $\sim
10^{-3} Z_\odot$ is not a true first generation object.

\section{Low Metallicity HII Regions}

We have also computed model HII regions for metallicities from three
times solar  down to $10^{-6} Z_\odot$\ (Panagia {\it et al.}~2004,
in preparation).  In Figure~1 the synthetic spectrum of an HII region
with metallicity $10^{-3} Z_\odot$ (third panel from the top) can be
compared to that of  an object with zero metallicity (top panel). The
two are very similar except for a few weak metal lines.  It is also
apparent that values of EW in excess of
1,000\AA\/ are possible already for objects with metallicity $\sim
10^{-3} Z_\odot$. This is particularly interesting given that
Ly-$\alpha$ emitters with large EW have been identified at z=5.6
(Rhoads \& Malhotra 2001).

  \begin{figure}
    \begin{center}
       \includegraphics[width=7cm]{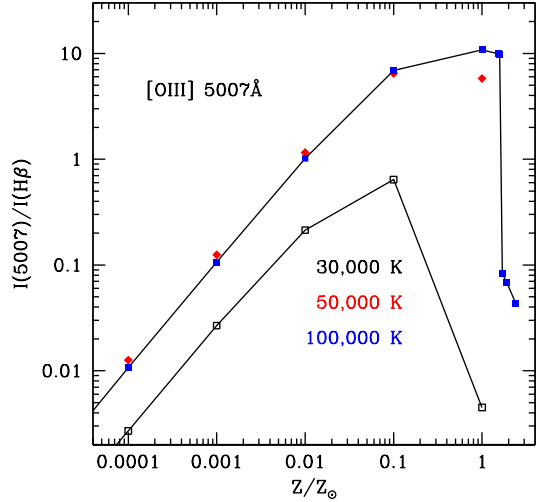}
   \end{center}
  \caption{{The ratio [OIII]$\lambda 5007$ / H$\beta$ is plotted as a
function of metallicity for three different stellar temperatures: 30,000K
(open squares and lower line), 50,000K (solid diamonds), and 100,000K
(solid squares and upper line).} 
} 
  \end{figure}

The metal lines are weak, but some of them can be used as metallicity
tracers. In Figure~2 the intensity ratio of the  [OIII]$\lambda 5007$
line to H$\beta$ is plotted for a range of stellar temperatures and
metallicities. We notice that for $Z < 10^{-2} Z_\odot$ this line
ratio traces metallicity linearly. Our reference value $Z = 10^{-3}$
corresponds to a ratio [OIII]/H$\beta$ = 0.1. The weak dependence on
stellar temperature makes sure that this ratio remains a good indicator
of metallicity also when one considers the integrated signal from a
population with a range of stellar masses.

Another difference between zero-metallicity and low-metallicity HII
regions lies in the possibility that the latter may contain dust. For
a $Z=10^{-3} Z_\odot$ HII region dust may absorb up to 30 \% of the
Ly-$\alpha$ line, resulting in roughly 15 \% of the energy being
emitted in the far IR (Panagia {\it et al.}~2004, in preparation).

\section{How to discover and characterize  Primordial HII Regions}

It is natural to wonder whether primordial HII regions will be
observable with the generation of telescopes currently on the drawing
boards. In this section we will focus mostly on the capabilities of the
James Webb Space Telescope ({\it JWST}; \eg Stiavelli \etal). Here we  
consider a starburst with
$\sim10^6$ M$_\odot$ in massive stars(which corresponds to a
Ly-$\alpha$  luminosity of $\sim10^{10}$ L$_\odot$) as  our reference
model.


  \begin{figure}
    \begin{center}
     
 \includegraphics[width=7.0cm]{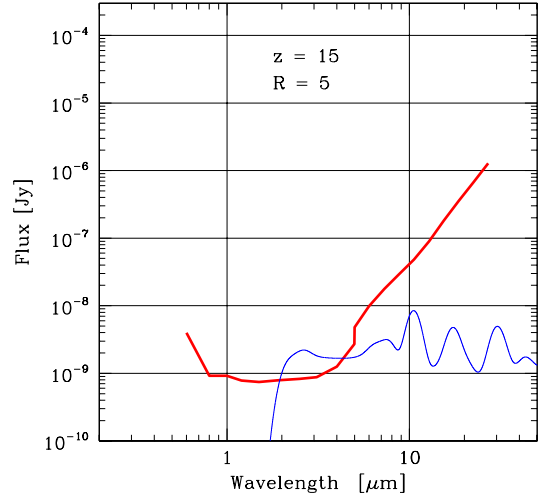}
   \end{center}
  \caption{{Synthetic spectral energy distribution of a Z=$10^{-3} Z_\odot$
starburst object at z=15 containing $10^6$ M$_\odot$ in massive stars
(thin line) compared to the imaging limit of $JWST$ at R=5 (thick line).
The $JWST$ sensitivity refers to $4\times10^5$ s exposures with S/N=10.}  
} 
  \end{figure}


  \begin{figure}
    \begin{center}
     
\includegraphics[width=7.0cm]{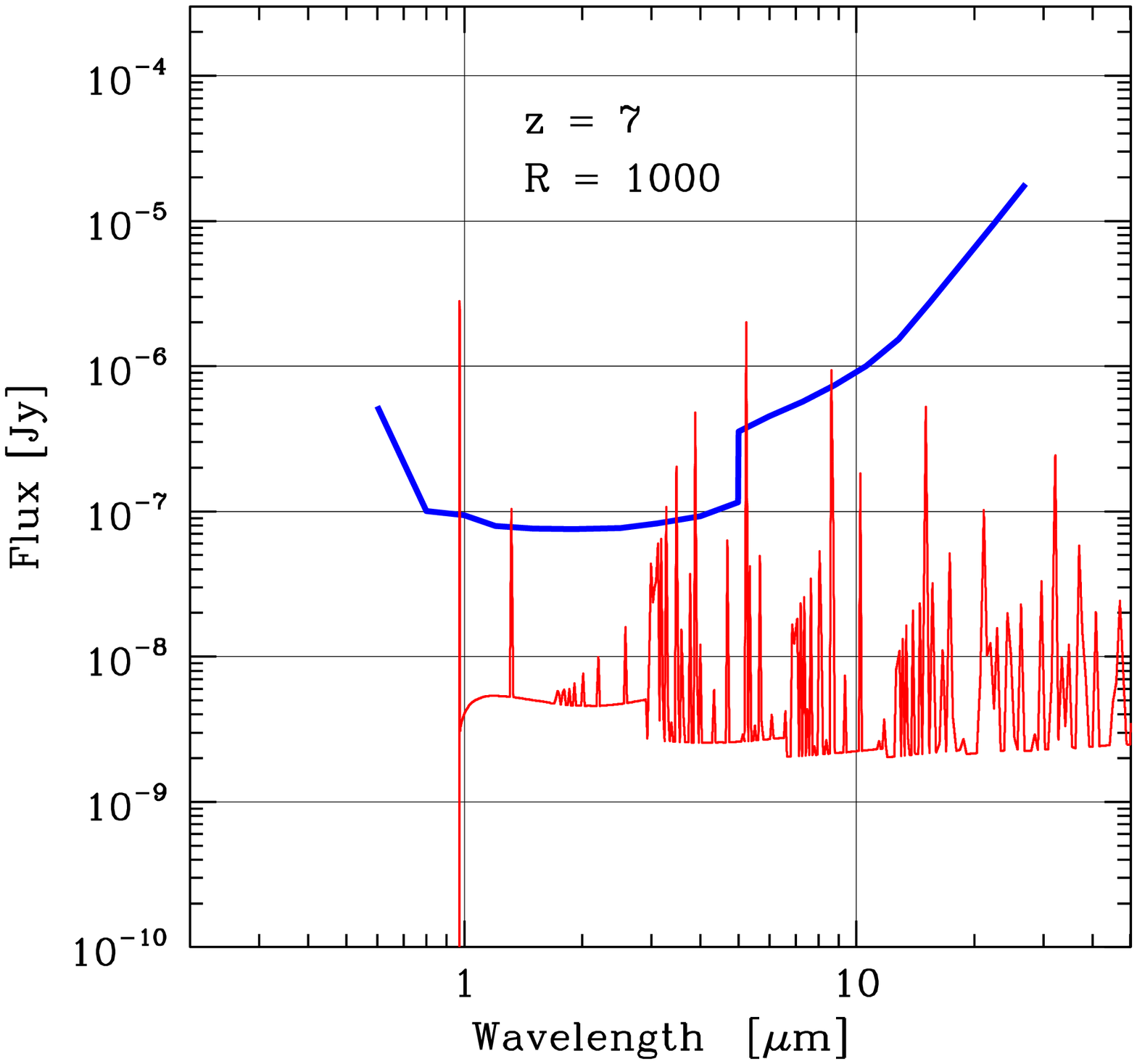}
    \end{center}
  \caption{{Synthetic spectrum of a Z=$10^{-3} Z_\odot$ starburst
object at z=7 containing $10^6$ M$_\odot$ in massive stars (thin line)
compared to the spectroscopic limit of $JWST$ at R=1000 (thick
line). The $JWST$ sensitivity refers to $4\times10^5s$  exposures
with S/N=10.}  
} 
  \end{figure}


The synthetic spectra, after allowance for IGM HI absorption of the
Ly-$\alpha$ radiation (\eg Miralda-Escud\'e \& Rees 1998,  Madau \&
Rees 2001) and convolution with suitable filter responses  are compared
to the ~{\it JWST} imaging sensitivity for $4\times10^5$s exposures in
Figure~3. It is clear that~{\it JWST} will be able to easily detect
such objects. Due to the high background from the ground,~{\it JWST}
will remain superior even to 30m ground based telescopes for these
applications.

The synthetic spectra can also be compared to the~{\it JWST}
spectroscopic sensitivity for $4\times10^5s$ exposures (see Figure~4). 
It appears that while the  Ly-$\alpha$ line can be detected up to
$z\simeq 15-20$, for our reference source only at relatively low
redshifts (z$\sim7$) can~{\it JWST} detect other diagnostics lines lines
such as HeII 1640\AA, and Balmer lines. Determining metallicities is
then limited to lower redshifts or to brighter sources.

  \begin{figure}
    \begin{center}
      \includegraphics[width=6.7cm]{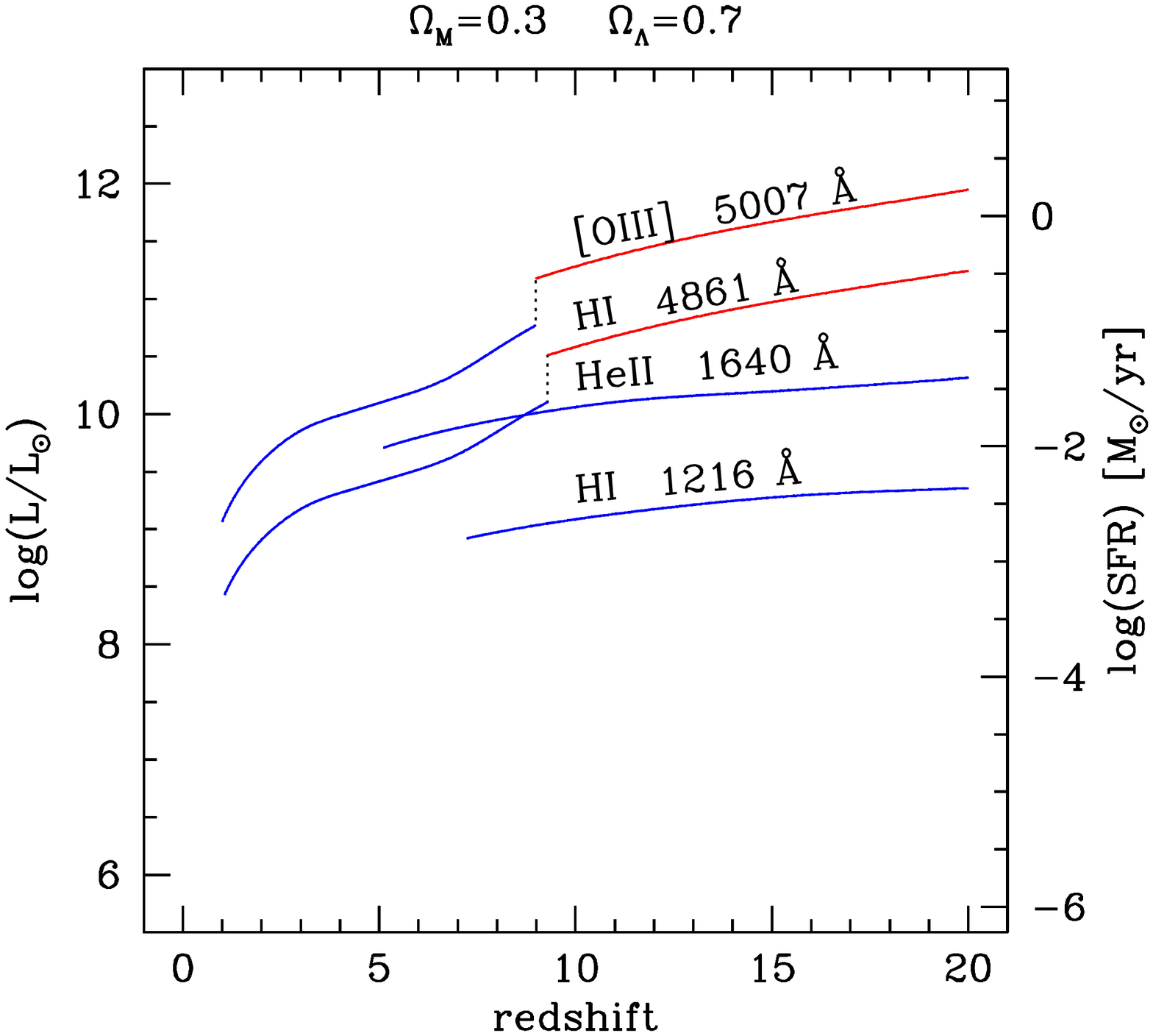}
    \end{center}
  \caption{{Limiting $total$ luminosity of the ionizing stars
  (left-hand scale) and top-heavy IMF star formation rate (right-hand
  scale) to detect various emission lines using  $JWST$ spectroscopy,
  with S/N=10 in integrations of 100 hours, as a function of the source
  redshift.} 
  } 
  \end{figure}
%
%

We can reverse the argument and ask ourselves what sources can ~{\it
JWST} detect and characterize with spectroscopic observations.  
Figure~5 displays, as a function of redshift,  the total luminosity of
a starburst  whose lines can be detected with a S/N=10 for a
$4\times10^5$s exposure time. The loci for Ly-$\alpha$, HeII 1640\AA,
H\/$\beta$, and [OIII] 5007\AA\/ are shown. It appears that Ly-$\alpha$
is readily detectable up to z$\simeq$20, the HeII 1640\AA\/ line may
also be detected up to high redshifts, whereas ``metallicity"
information,  {\it i.e.} the intensity ratio I([OIII])/I(H$\beta$), can
be obtained at high z only for sources that are $\sim30$ times more
massive or that are magnified $\sim30$ times by gravitational
lensing.  

\section {Primordial Supernovae}

Even if {\it JWST} cannot detect individual massive Population III
stars, supernova explosions may come to the rescue. In the local
Universe supernovae (SNe) can be as bright as an entire galaxy ({\it
e.g.}, Type Ia supernovae (SNIa) at maximum light have
M$_B(SNIa)\simeq$-19.5) and are detectable up to large distances.
However, SNIa, originating from moderate mass stars, are not  expected
to occur during the first 1 billion years after Big Bang. In addition,
SNIa are efficient emitters only at rest frame wavelengths longer than
2600A, which makes them hard to detect at high redshifts (Panagia
2003a,b). Type II supernovae (SNII) are much more efficient UV emitters
but only rarely they are as bright as a SNIa. As a consequence, they
will barely be detected at redshifts higher than 10, or, if they are
exceptionally bright (\`a la SN 1979C or SN 1998S) they  would be rare
events (Panagia 2003a,b).

On the other hand, massive population III stars are much more massive
than Pop II or Pop I stars, and the resulting supernovae may have
properties very different from those of local Universe SNe. Heger et al
(2001) have considered the fate of massive stars in conditions of zero
metallicity and have found that  for stellar progenitors with masses in
the range 140-260 M$_\odot$  the SN explosions would be caused by a
pair-production instability and would be 3 to 100 times more powerful
than core-collapse (Type II and Type Ib/c) SNe, so that a Pop~III SN at
a redshift of z = 20 could attain a peak flux  of about 100 nJy at 5
$\mu m$. Such a high flux would be easily detectable with {\it JWST}
observations made with an integration time of a few hours.

Next point to consider is: ``Do Pop~III SNe occur frequently enough to
be found in a systematic search?" For a standard cosmology
($\Omega_\Lambda = 0.7$, $\Omega_m = 0.3$, H$_0=65 ~km~s^{-1}Mpc^{-1}$,
$\Omega_b=0.047$),  and assuming that at z=20 a fraction $10^{-6}$ of
all baryons goes into stars of 250 M$_\odot$,  Heger {\it et al.}
(2001) predict an overall rate of 0.16 events per second over the
entire sky, or about $3.9\times10^{-6}$ events per second per square
degree. Since for these primordial SNe the first peak of the light
curve lasts about a month, about a dozen of these SNe per square degree
should be at the peak of their light curves at any time. Therefore, by
monitoring about 100 NIRCam fields with integration times of about
10,000 seconds at regular intervals (every few months) for a year
should lead to the discovery of three of these primordial supernovae.
We conclude that, with a significant investment of observing time (a
total of 4,000,000 seconds) and with a little help from Mother Nature
(to endorse our theorists views), ~{\it JWST} will be able to detect
{\it individually} the very first stars and light sources in the Universe.


\begin{chapthebibliography}{}
\bibitem[2001]{}   Baraffe I., Heger A., Woosley S. E.,  2001, ApJ, 550, 890 
\bibitem[2001]{}   Bromm V., Ferrara A., Coppi P. S., Larson R. B., 2001, MNRAS, 328, 969
\bibitem[2001]{}   Bromm V., Kudritzki R. P., Loeb, A., 2001, ApJ, 552, 464
\bibitem[2001]{}   Heger A., Woosley S. E., Baraffe I., Abel T., 2001, astro-ph 0112059
\bibitem[2001]{}   Madau P., Rees M. J., 2001, ApJ, 551, L27
\bibitem[2001]{}   Marigo P., Girardi L., Chiosi C., Wood P. R. 2001, A\&A 371, 252
\bibitem[1998]{}   Miralda-Escud\'e J., Rees M. J., 1998, ApJ, 497, 21
\bibitem[2002]{}   Oh S. P., Haiman Z., 2002, ApJ, 569, 558
\bibitem[1983]{}   Oliva E., Panagia N., 1983, Ap\&SS, 94, 437
\bibitem[2003]{}   Panagia N., 2003a, in ``Supernovae and Gamma-Ray Bursters", 
			ed. K. W. Weiler (Berlin: Springer-Verlag), p. 113-144
\bibitem[2003]{}   Panagia N., 2003b,  STScI Newsletter, Vol. 20, Issue 4, p. 12
\bibitem[2001]{}   Rhoads, J. E., Malhotra S., 2001, ApJ, 563, L5
\bibitem[2001]{}   Stiavelli, M., et al. 2004, ``JWST Primer",
			(Baltimore: STScI)
\bibitem[2000]{}   Tumlinson J., Shull J. M., 2000, ApJ, 528, L65
\bibitem[2001]{}   Tumlinson J., Giroux M. L., Shull J. M., 2001, ApJ, 550, L1
\end{chapthebibliography}
\end{document}